\documentclass[aps,prl,twocolumn,showpacs,superscriptaddress]{revtex4}
\usepackage{graphicx} % Include figure files
\usepackage{dcolumn}  % Align table columns on decimal point
\usepackage{bm}       % bold math
\usepackage{amsmath}
\usepackage{epsfig}

\def\ii{{\rm i}}  \def\ee{{\rm e}}  \def\rb{{\bf r}}  \def\Rb{{\bf R}}
  \def\jb{{\bf j}}    \def\Eb{{\bf E}}
\def\vb{{\bf v}}  
\def\nt{{\hat{\bf n}}}

\def\zt{{\hat{\bf z}}}

\begin{document}

%\title{Electron energy loss spectroscopy as a probe of the photonic local density of states}
\title{Probing the photonic local density of states with electron energy loss spectroscopy}
\author{F. J. Garc\'{\i}a de Abajo}
\affiliation{Instituto de \'Optica - CSIC, Serrano 121, 28006
Madrid, Spain}

\author{M. Kociak}
\affiliation{Laboratoire de Physique des Solides, CNRS, Universit\'e
Paris Sud XI, F 91405 Orsay, France}

\date{\today}

\begin{abstract}
Electron energy-loss spectroscopy (EELS) performed in transmission
electron microscopes is shown to directly render the photonic local
density of states (LDOS) with unprecedented spatial resolution,
currently below the nanometer. Two special cases are discussed in
detail: (i) 2D photonic structures with the electrons moving along
the translational axis of symmetry and (ii) quasi-planar plasmonic
structures under normal incidence. Nanophotonics in general and
plasmonics in particular should benefit from these results
connecting the unmatched spatial resolution of EELS with its ability
to probe basic optical properties like the photonic LDOS.
\end{abstract}
 \maketitle

While a plethora of nanophotonic structures are currently being
devised for diverse applications like achieving single molecule
sensitivity in biosensing \cite{BCN05} or molding the flow of light
over nanoscale distances for signal processing \cite{BDE03}, no
optical characterization technique exists that can render
spectroscopic details with truly nanometer spatial resolution. The
need for that kind of technique is particularly acute in nanometric
plasmonic designs that benefit from sharp edges and metallic
surfaces in close proximity to yield large enhancements of the
electromagnetic field.

Scanning transmission electron microscopes (STEM) can plausibly
cover this gap, as they perform electron energy loss spectroscopy
(EELS) with increasingly improved energy resolution that is quickly
approaching the width of plasmon excitations in noble metals
\cite{paper125} and with spatial resolution well below the nanometer
\cite{BDK02}. The connection between EELS and photonics can be
readily established when low-energy losses in the sub-eV to a few eV
range are considered, compatible with typical photon energies in
photonic devices. A formidable amount of information is available in
the literature for this so-called valence EELS, including for
instance studies of single nanoparticles of various shapes
\cite{HM1985,paper125}, interacting nanoparticles \cite{UCT92}, thin
films \cite{CS1975}, composite metamaterials \cite{HW91}, and carbon
nanostructures \cite{STK02}. Many of these reports are relevant to
current nanophotonics research, in which the optical response of
nanoparticles, nanoparticle assemblies, and patterned nanostructures
plays a central role. In this context, EELS has been recently
demonstrated to image plasmon modes with spatial resolution better
than a hundredth of the wavelength in triangular nanoprisms
\cite{paper125}. However, despite significant progress from the
theoretical side \cite{RZA00}, no synthetic and universal picture
has emerged to explain the spatial modulation of EELS measurements
on arbitrary nanostructures.

In this Letter, we show that EELS provides direct information on the
photonic local density of states (LDOS), and thus it constitutes a
suitable tool for truly nanometric characterization of photonic
nanostructures. A rigorous derivation of this statement is offered,
illustrated by numerical examples for both translationally-invariant
geometries and planar structures. Our results allow directly
interpreting EELS data in terms of local photonic properties that
encompass the full display of optical phenomena exhibited by
nanostructures, ranging from localized and propagating plasmons in
patterned metallic surfaces \cite{UCT92} to band gaps in dielectric
photonic crystals \cite{paper080}.

{\it Green tensor and LDOS.--} The optical response of a
nanostructure is fully captured in its electric Green tensor and its
LDOS \cite{note3}. In particular, the electric field produced by an
external current density $\jb(\rb,\omega)$ in an inhomogeneous
medium of permittivity $\epsilon(\rb,\omega)$ can be written in
frequency space $\omega$ as
\begin{eqnarray}
\Eb(\rb,\omega)=-4\pi\ii\omega \int d\rb' G(\rb,\rb',\omega)
\jb(\rb',\omega) \label{eq6}
\end{eqnarray}
in terms of $G$, the electric Green tensor of Maxwell's equations in
Gaussian units, satisfying
\begin{eqnarray}
&&\nabla\times\nabla\times G(\rb,\rb',\omega)-(\omega^2/c^2)
\epsilon(\rb,\omega) G(\rb,\rb',\omega) \nonumber \\ && =
\frac{-1}{c^2}\delta(\rb-\rb') \label{defgreen}
\end{eqnarray}
and vanishing far away from the sources.

We then define the LDOS projected along unit a vector $\nt$ as
\cite{FMM05}
\begin{eqnarray} \rho_\nt(\rb,\omega)=\frac{-2\omega}{\pi} {\rm Im}\{\nt\cdot
G(\rb,\rb,\omega)\cdot\nt\}. \label{eq2}
\end{eqnarray}
In free space, the uniform LDOS is known from black-body theory:
\begin{eqnarray}
\rho^0_\nt(\rb,\omega)=\omega^2/3\pi^2c^3. \label{eq4}
\end{eqnarray}
Similar to its electron counterpart in solid state physics, the
photonic LDOS equals the combined local intensity of all eigenmodes
of the system under consideration, provided they are well defined
(e.g., in the absence of lossy media) \cite{CGW01}. An alternative
interpretation, which holds even in the presence of lossy materials
\cite{DMC04}, comes from the realization that $(4\pi^2\omega
D^2/\hbar)\,\rho$ is the decay rate for an excitation dipole
strength $D$ \cite{BL96,FMM04}. Finally, we point out that a
complete definition of the LDOS should include a magnetic part
\cite{JCM03}, which is however uncoupled to our fast electrons.

{\it Energy loss probability.--} The energy loss suffered by a fast
electron passing near an inhomogeneous sample and moving with
constant velocity $\vb$ along a straight line trajectory
$\rb=\rb_e(t)$ can be related to the force exerted by the induced
electric field $\Eb^{\rm ind}$ acting back on the electron as
\cite{R1957}
   \begin{eqnarray}
      \Delta E = e \int dt \; \vb\cdot\Eb^{\rm ind}[\rb_e(t),t]
               = \int_0^\infty \hbar\omega\, d\omega \;
               \Gamma(\omega),
   \label{eq9}
   \end{eqnarray}
where the $-e$ electron charge has been included (i.e., $\Delta
E>0$) and
   \begin{eqnarray}
      \Gamma(\omega) = \frac{e}{\pi\hbar\omega} \int dt \,
                       {\rm Re}\left\{ \ee^{-\ii\omega t}
              \vb\cdot\Eb^{\rm ind}[\rb_e(t),\omega]\right\}
   \label{eq10}
   \end{eqnarray}
is the loss probability. The Fourier transform
   \begin{eqnarray}
      \Eb^{\rm ind}(\rb,t) = \int \frac{d\omega}{2\pi}
                             \ee^{-\ii\omega t}
                             \Eb^{\rm ind}(\rb,\omega)
   \label{eq11}
   \end{eqnarray}
has been introduced and the property $\Eb^{\rm
ind}(\rb,\omega)=[\Eb^{\rm ind}(\rb,-\omega)]^*$ has been used.

The external current density corresponding to the moving electron
is now given by
   \begin{eqnarray}
      \jb(\rb,\omega)=-e\vb \int dt \ee^{\ii\omega t}
      \delta[\rb-\rb_e(t)].
   \label{eq12}
   \end{eqnarray}
Assuming without loss of generality that the velocity vector is
directed along the positive $z$ axis and using the notation
$\rb=(\Rb,z)$, with $\Rb=(x,y)$, the current density reduces to
   \begin{eqnarray}
      \jb(\rb,\omega)=-e\delta(\Rb-\Rb_0) \ee^{\ii\omega z/v}\,\zt,
   \label{eq13}
   \end{eqnarray}
where $\Rb_0=(x_0,y_0)$ is the 2D impact parameter of the electron
trajectory relative to the $z$ axis. Inserting Eq.\ (\ref{eq13})
into Eq.\ (\ref{eq6}), and this in turn into Eq.\ (\ref{eq10}), we
find
   \begin{eqnarray}
      && \Gamma(\Rb_0,\omega) = \label{eq14} \\
      && -\frac{4e^2v^2}{\hbar} \int dt\,dt' \,
                       {\rm Im}\left\{\ee^{\ii\omega (t'-t)}
              G^{\rm ind}_{zz}[\rb_e(t),\rb_e(t'),\omega]\right\},
   \nonumber
   \end{eqnarray}
where $G_{zz}=\zt\cdot G\cdot\zt$, the dependence of the loss
probability $\Gamma$ on $\Rb_0$ is shown explicitly, and $G^{\rm
ind}$ denotes the induced Green tensor obtained from $G$ by
subtracting the free-space Green tensor.

{\it Relation between EELS and LDOS.--} Noticing that $z_e(t)=vt$,
the time integrals of Eq.\ (\ref{eq14}) yield the Fourier transform
of the induced Green tensor with respect to $z$ and $z'$, $G^{\rm
ind}_{zz}(\Rb,\Rb',q,-q',\omega)$, in terms of which $\Gamma$
becomes
   \begin{eqnarray}
      \Gamma(\Rb_0,\omega) &=&
      -\frac{4e^2}{\hbar} \, {\rm Im} \{G^{\rm ind}_{zz}(\Rb_0,\Rb_0,q,-q,\omega)\},
      \nonumber \\ &=& \frac{2\pi e^2}{\hbar\omega} \rho_\zt(\Rb_0,q,\omega),
   \label{eq14bis}
   \end{eqnarray}
where $q=\omega/v$ and we have defined
\begin{eqnarray}
\rho_\nt(\Rb,q,\omega)=\frac{-2\omega}{\pi} {\rm Im}\{\nt\cdot
G(\Rb,\Rb,q,-q,\omega)\cdot\nt\} \label{eq2bis}
\end{eqnarray}
as a generalized density of states that is local in real space along
the $\Rb$ directions and local in momentum space along the remaining
$z$ direction, parallel to the electron velocity vector.

The value of $q=\omega/v$ reflects conservation of energy and
momentum in the transfer of excitations of frequency $\omega$ and
momentum $q$ from the electron to the sample. It is interesting to
note that for an electron moving in an infinite vacuum one has
   \begin{eqnarray}
      \rho_{\zt}(\Rb,q,\omega)=L\,\frac{\omega}{2\pi c^2}
      \theta(\omega/c-q),
   \label{eq18}
   \end{eqnarray}
which is always zero for subliminal electrons moving with velocity
$v<c$. Here, $L$ is the {\it quantization} length along $z$. As
expected, the vacuum density of states does not contribute to the
EELS signal, and this allows dropping the superscript {\it ind} from
Eq.\ (\ref{eq14bis}), as it has been implicitly assumed already when
writing Eq.\ (\ref{eq2bis}). Incidentally, the Cherenkov effect is
deduced from $\rho_{\zt}(\Rb,q,\omega)=L\,(\omega/2\pi c^2)
(1-k^2/q^2\epsilon) \theta(\epsilon\omega^2/c-q^2)$, valid for a
homogeneous dielectric of real permittivity $\epsilon$.

The proposed formalism, yields exactly the same results as any other
local, retarded theory, but it provides a new paradigm for
understanding EELS as connected to a local quantity: the LDOS. This
is in contrast to the common view of EELS as a tool capable of
retrieving local electronic properties hidden in the permittivity,
which is in turn involved in the non-local response of inhomogeneous
structures probed by the electrons and has resulted in endless
discussions regarding how to eliminate delocalization.

{\it 2D systems possessing translational invariance.--} If the
sample under consideration is translationally-invariant along $z$,
the Green tensor $G(\rb,\rb',\omega)$ depends on $z$ and $z'$ only
via $z-z'$, and thus one can write
   \begin{eqnarray}
      G(\rb,\rb',\omega)=\int \frac{dq}{2\pi} \tilde{G}(\Rb,\Rb',q,\omega) \ee^{\ii q
      (z-z')}.
   \label{eq15}
   \end{eqnarray}
where $\tilde{G}(\Rb,\Rb',q,\omega)=(1/L)G(\Rb,\Rb',q,-q,\omega)$.
Accordingly, the local density of states can be decomposed into
momenta components $q$ along the $z$ axis,
   \begin{eqnarray}
      \rho_\nt(\rb,\omega)=\int \frac{dq}{2\pi} \tilde{\rho}_\nt(\Rb,q,\omega),
   \label{eq16}
   \end{eqnarray}
where
\begin{eqnarray}
\tilde{\rho}_\nt(\Rb,q,\omega)&=&\frac{-2\omega}{\pi} {\rm Im}\{\nt\cdot
\tilde{G}(\Rb,\Rb,q,\omega)\cdot\nt\} \nonumber \\ &=& \frac{1}{L}
\rho_\nt(\Rb,q,\omega). \label{eq17}
\end{eqnarray}
Finally, combining Eqs.\ (\ref{eq14bis}) and (\ref{eq17}), one finds
a relation between the loss probability per unit of path length and
the LDOS,
   \begin{eqnarray}
      \frac{\Gamma(\Rb_0,\omega)}{L} = \frac{2\pi e^2}{\hbar\omega}
      \tilde{\rho}_\zt(\Rb_0,q,\omega),
   \label{eq19}
   \end{eqnarray}
with $q=\omega/v$.

Eq. (\ref{eq19}) provides a solid link between LDOS and EELS in a
wide class of geometries that include aloof trajectories in
semi-infinite surfaces and thin films. Interestingly, only
$q>\omega/c$ values are probed, lying outside the light cone, and
therefore difficult to study via optical techniques. Trapped modes
such as surface-plasmon polaritons lie in that region and are a
natural target for application of our results. Besides, the present
study can be directly applied to cathodoluminescence (CL) in
all-dielectric structures, in which energy loss and CL emission
probabilities are identical.

A connection between the  photonic density of states in the momentum
space and EELS has been previously reported for electrons moving
parallel to pores in 2D self-assembled alumina photonic crystals
\cite{paper080}. However, the above derivation is the first prove to
our knowledge that a formal relation exists between LDOS and EELS.

\begin{figure}[ht]
\includegraphics[width=80mm,angle=0,clip]{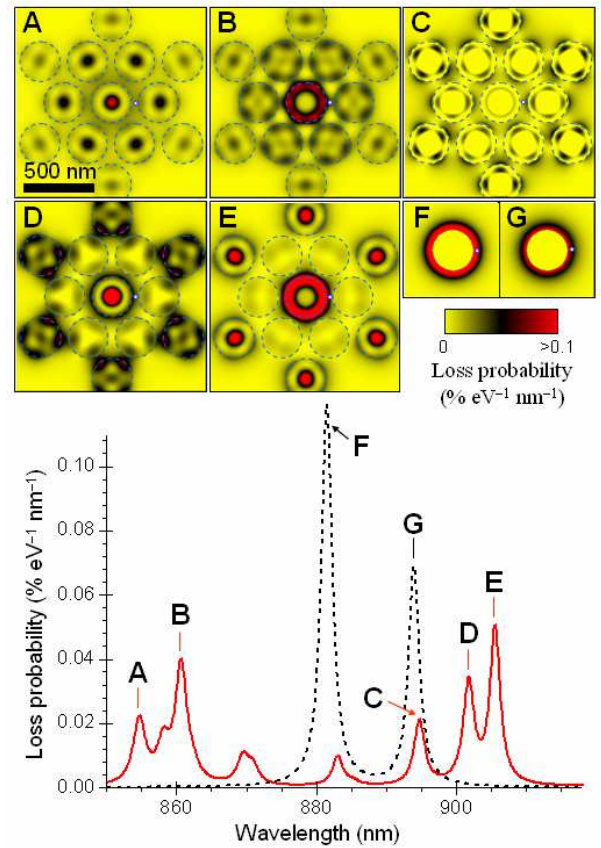}
\caption{EELS probability for 200-keV electrons moving parallel to
an array of 13 Si nanowires of 300 nm in diameter (solid curve), as
compared to an isolated nanowire (broken curve). The nanowires are
arranged in a hexagonal lattice with a period of 350 nm. The
contour-plot insets show the impact-parameter dependence of the EELS
probability for selected energy losses (labeled A-G), with the
position of the electron beam considered in the curves indicated by
an open symbol and the cylinder contours shown by dashed curves.}
\label{2D}
\end{figure}

For illustration, we offer in Fig.\ \ref{2D} the EELS probability
for electrons moving inside a finite hexagonal 2D crystal of aligned
Si nanowires, calculated with the boundary element method (BEM)
\cite{paper070}. The photon wavelength range under consideration
includes two Mie modes of the isolated wire (broken curve), and
significant hybridization between neighboring wires takes place in
the array. The loss probability varies relatively smoothly with
impact parameter, a behavior which was expected in the LDOS for
photon wavelengths relatively large compared to the cylinders
diameter. Interestingly, the loss probability takes significant
values in the intersticial regions, several tens of nanometers away
from the Si. Finite structures like that considered in Fig.\
\ref{2D} exhibit a colorful evolution of modes, the analysis of
which can be useful for instance in the design of microlaser
cavities. In infinite crystals, the loss probability exhibits Van
Hove singularities \cite{V1953}, which follow rigorously those of
the LDOS in translationally-invariant systems according to Eq.\
(\ref{eq19}).

\begin{figure*}[ht]
\includegraphics[width=155mm,angle=0,clip]{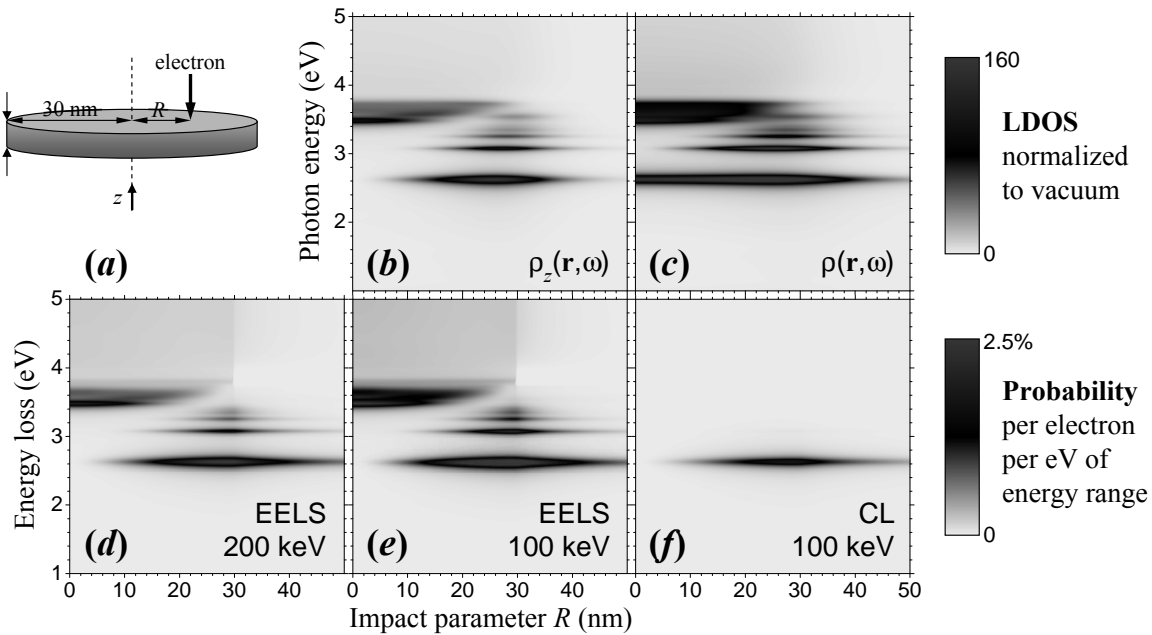}
\caption{Relation between EELS and LDOS in planar geometries. {\bf
(a)} We consider an Ag disk of height 10 nm and radius 30 nm. The
electrons move along $z$, perpendicular to the disk. The EELS
probability for 200-keV {\bf (d)} and 100-keV {\bf (e)} electrons
mimics closely the $z$-projected LDOS in a plane 10 nm above the
disk {\bf (b)}, and less closely the unprojected LDOS
($\rho=\rho_x+\rho_y+\rho_z$) {\bf (c)}. The cathodoluminescence
emission only picks up part of the inelastic signal {\bf (f)}.}
\label{planar}
\end{figure*}

{\it Planar geometries.--} As microchip features continue to shrink,
lithographically-patterned metal structures are becoming natural
candidates to replace current electronic microcircuits. The new
structures will operate at frequencies above the THz, rather than
GHz, and will carry electric signals strongly mixed with the
electromagnetic fields that they generate in what is known as
surface plasmons. The optical properties of metallic planar
structures are routinely obtained using scanning near-field optical
microscopy (SNOM), although the lateral resolution of this technique
can hardly reach 50 nm. Here again EELS renders much higher lateral
resolution (down to the nm) and permits obtaining information
directly related to the LDOS. Besides the formal relation between
EELS and the momentum-resolved LDOS expressed in Eq.\
(\ref{eq2bis}), we offer in Fig.\ \ref{planar} a more detailed
comparison between EELS and $\rho(\rb,\omega)$ (local in all spatial
directions) for an Ag disk calculated with BEM \cite{paper070}. This
figure proves how the EELS probability can mimic quite closely
$\rho_z(\rb,\omega)$, with $z$ perpendicular to the planar structure
and parallel to the electron velocity. This resemblance holds for
different accelerating voltages, making the interpretation robust
with respect to experimental details. Fig.\ \ref{planar} provides a
solid example supporting the use of EELS to measure plasmon
intensities with unprecedented lateral resolution. A first
demonstration of such type of measurements has been recently
reported \cite{paper125}.

We thus conclude that the energy loss probability is directly
related to the local density of states in arbitrary systems, where
we understand locality in real space for the directions
perpendicular to the electron trajectory and in momentum space along
the direction of the electron velocity vector. In 2D systems and for
electrons moving along the direction of translational symmetry, the
loss probability is exactly proportional to the photonic local
density of states projected on the trajectory and decomposed into
parallel momentum transfers $q$. Numerical examples have been
presented showing a similar relation between LDOS outside planar
metallic disks and EELS spectra for electrons traversing them
perpendicularly. Our results provide a solid foundation for the use
of EELS performed in STEMs to directly probe photonic properties of
nanostructures.

\begin{acknowledgments}
The authors thank R. C. McPhedran and J. J. Greffet for helpful and
enjoyable discussions. This work was supported by the Spanish MEC
(FIS2004-06490-C03-02) and by the EU (STRP-016881-SPANS).
\end{acknowledgments}

%\bibliography{../bibtex/refs}
%\bibliographystyle{prsty}

\end{document}